\pdfoutput=1
\documentclass[twocolumn,pre,amsmath,amssymb]{revtex4-1}
\usepackage{graphicx}
\usepackage{amsmath}
\usepackage{bm}
\usepackage{xcolor}
\usepackage{dcolumn}
\usepackage{ulem}
\usepackage{hhline}
\usepackage{float}

\begin{document}

\title{Models of infiltration into homogeneous and fractal porous media with localized sources}

\author{Fabio D. A. Aar\~ao Reis}
\email{reis@if.uff.br}
\affiliation{Instituto de F\'\i sica, Universidade Federal Fluminense,\\
Avenida Litor\^anea s/n, 24210-340 Niter\'oi RJ, Brazil}
\author{ Vaughan R. Voller}
\email{volle001@umn.edu}
\affiliation{Department of Civil, Environmental, and Geo- Engineering,\\
Saint Anthony Falls Laboratory, University of Minnesota,\\
500 Pillsbury Drive SE, Minneapolis, MN 55455, USA }
\date{\today}

\begin{abstract}

We study a random walk infiltration (RWI) model, in homogeneous and in fractal media,
with small localized sources at their boundaries.
In this model, particles released at a source, maintained at a constant density
value, execute unbiased random walks over a lattice; a model that represents  solute
infiltration by diffusion into a medium in contact with a reservoir of fixed concentration.
A scaling approach shows that the infiltrated length, area, or volume evolves in time as the
number of distinct sites visited by a single random walker in the same medium.
This is consistent with numerical simulations of the lattice model and exact and numerical
solutions of the corresponding diffusion equation.
In a Sierpinski carpet, the infiltrated area is expected to evolve as $t^{D_F/D_W}$
(Alexander-Orbach relation), where $D_F$ is the fractal dimension of the medium and $D_W$ is
the random walk dimension; the numerical integration of the diffusion equation supports this
result with very good accuracy and improves results of lattice random walk simulations.
In a Menger sponge in which $D_F>D_W$ (i.e., a fractal with a dimension close to $3$), a linear time
increase of the infiltrated volume is theoretically predicted and confirmed numerically.
Thus, no evidence of fractality can be observed in measurements of infiltrated volumes or masses in
media where random walks are not recurrent, although the tracer diffusion is anomalous.
We compare our findings with results for a fluid infiltration model in which the pressure head
is constant at the source and the front displacement is driven by the local gradient of that head.
Exact solutions in two and three dimensions and numerical results in a carpet show that
this type of fluid infiltration is in the same universality class of RWI, with an equivalence
between the head and the particle concentration.
These results set a relation between different infiltration processes with localized sources
and the recurrence properties of random walks in the same media.

\end{abstract}

\maketitle

\section{Introduction}
\label{introduction}

\begin{figure}[!ht]
\center
\includegraphics[clip,width=.45\textwidth,angle=0]{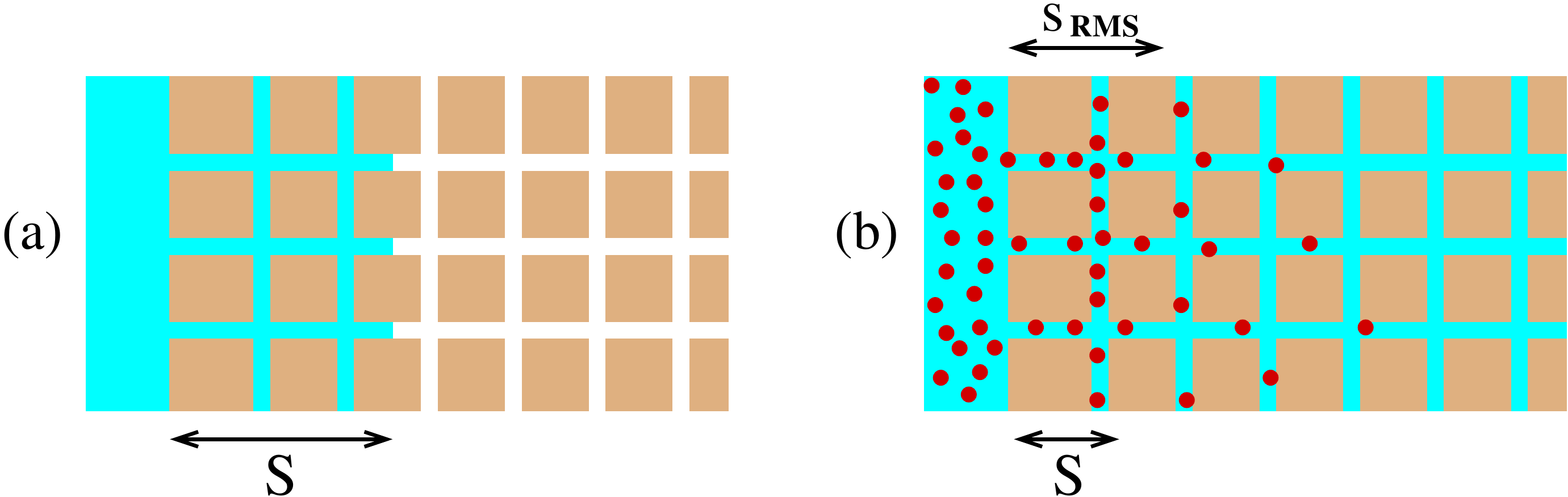}
\caption{
(a) Schematic of fluid (blue) infiltration from a high pressure reservoir at the left into
a porous media (solid in tan color, non-filled pores in white).
(b) Infiltration of a solute (red) from a reservoir at the left into a porous media initially
without that solute.
The characteristic lengths $s_{RMS}$ and $S$ are indicated for one-dimensional infiltration.
}
\label{infiltratedmedia}
\end{figure}

An infiltration process takes place when the external surface of an initially dry porous
medium is in contact with water source at a fixed pressure head.
In this circumstance, assuming that the Darcy law holds, the gradient of the pressure head will move
moisture into and through the pore spaces, as schematically shown in Fig. \ref{infiltratedmedia}(a).
If the domain geometry constrains the moisture to a one-dimensional horizontal flow, we can invoke a
Green-Ampt \cite{GreenAmpt} approximation and assume a well defined moving front $S\left( t\right)$
between filled and empty pore spaces.
Under this scenario, it would be expected that the filled volume $F\left( t\right)$ will increase in
time as
\begin{equation}
F \sim t^n ,
\label{defn}
\end{equation}
with the infiltration exponent $n=1/2$ \cite{campbellnorman, Peng, Chen}.
However, in many systems a different exponent is measured
\cite{logsdon,lockington,Kuntz,wilson,elabd2004,gerasimov,Voller_wrr2,elabd2015,berli2017},
which characterizes some type of anomalous transport (subdiffusive if $n<1/2$,
superdiffusive if $n>1/2$).

A related process occurs when the porous medium is filled with a static fluid and the external
surface is in contact with a reservoir of the same fluid but with a different concentration
of a solute.
This solute then infiltrates into or out of that medium, depending on the concentration difference,
with the solute molecules moving diffusively; see Fig. \ref{infiltratedmedia}(b).
A model for this process in a two-dimensional lattice was proposed by Sapoval, Rosso, and Gouyet
\cite{sapovalJPL1985}, and it was recently extended to fractal lattices \cite{Reis2016}:
the solute is represented by particles that execute random walks in the lattice and that have
a constant concentration at an external boundary.
Here, this is termed random walk infiltration (RWI).
Ref. \protect\cite{Reis2016} used a scaling approach and numerical simulations to show the
relation between the exponent $n$, the fractal dimensions of the infiltrated medium ($D_F$) and
of the infiltration boundary ($D_B$), and the random walk dimension in that medium ($D_W$).

In this work, we study RWI in homogeneous and fractal media with a small source at an external
boundary and discuss the close relation between this problem and fluid infiltration in the same media.
For two- and three-dimensional media, the time evolutions of the infiltrated area and volume
are obtained from the solution of the diffusion equation and confirmed by numerical simulations.
A scaling approach is then used to show that the infiltration has the same time evolution of the
number of distinct sites visited by a single random walker in the infiltrated lattice.
This leads to estimates of the infiltration exponent in fractal media, including the prediction
of an apparently normal infiltration in fractals where $D_F>D_W$, as illustrated for a Menger sponge.
In fluid infiltration, where the sharp fluid front moves with a velocity proportional
to the pressure gradient, the analytical form of the time evolution of the infiltrated area 
or volume in two- and three-dimensional media exactly matches that seen in  geometrically 
equivalent RWI processes.
Additional  numerical simulations also reveal similarities for fluid infiltration and RWI
in Sierpinski obstacle carpets. Together these results confirm the
scaling equivalence of fluid infiltration with RWI, and establish a connection between
different infiltration models with localized sources and the recurrence properties of random walks.

Despite the simple features of the fractal media studied here, we recall that they were used in
recent laboratory studies of anomalous transport.
For instance, infiltration of a fluid in horizontal Hele-Shaw cells with obstacles distributed
as in the Sierpinski carpets was studied in Ref. \protect\cite{Voller_wrr2} and Darcy-like flow
in media formed by sequences of Menger sponges was studied in Ref. \protect\cite{balankin2016}.
In both cases, good agreement with simulation results or with scaling approaches were obtained.
Hence it is expected that the scope of the work herein will benefit our understanding of
transport in porous materials of current geological and technological interest.

This work is organized as follows.
In Sec. \ref{model}, the RWI model and the basic quantities to be measured are presented,
and the example of infiltration in two dimensions with a linear boundary is discussed.
In Sec. \ref{localized}, we study RWI with localized sources in two- and three-dimensional
homogeneous media and in fractal lattices embedded in two and three dimensions, and establish
the connection with recurrence properties of random walks.
In Sec. \ref{fluid}, we study fluid infiltration in homogneous media and in a fractal.
In Sec. \ref{conclusion}, we summarize our results and present our conclusions.

\section{Random walk infiltration}
\label{model}

\subsection{Model definition}
\label{definition}

RWI is defined as a process in which particles are released from a static source
with constant concentration and in which those particles execute random walks in a lattice with
excluded volume interaction.
We denote as $a$ the edge of a lattice site, which can be occupied by at most one particle at a time.
In the time unit $\tau$, each particle attempts to hop to a randomly chosen nearest neighbor (NN) site; 
the hop is allowed only if that site is empty, otherwise the particle does not move.
The diffusion coefficient of a free particle in a free lattice of dimension $d$ is
\begin{equation}
D=\frac{a^2}{2d\tau} .
\label{defD}
\end{equation}

\begin{figure}[!ht]
\center
\includegraphics[clip,width=.45\textwidth,angle=0]{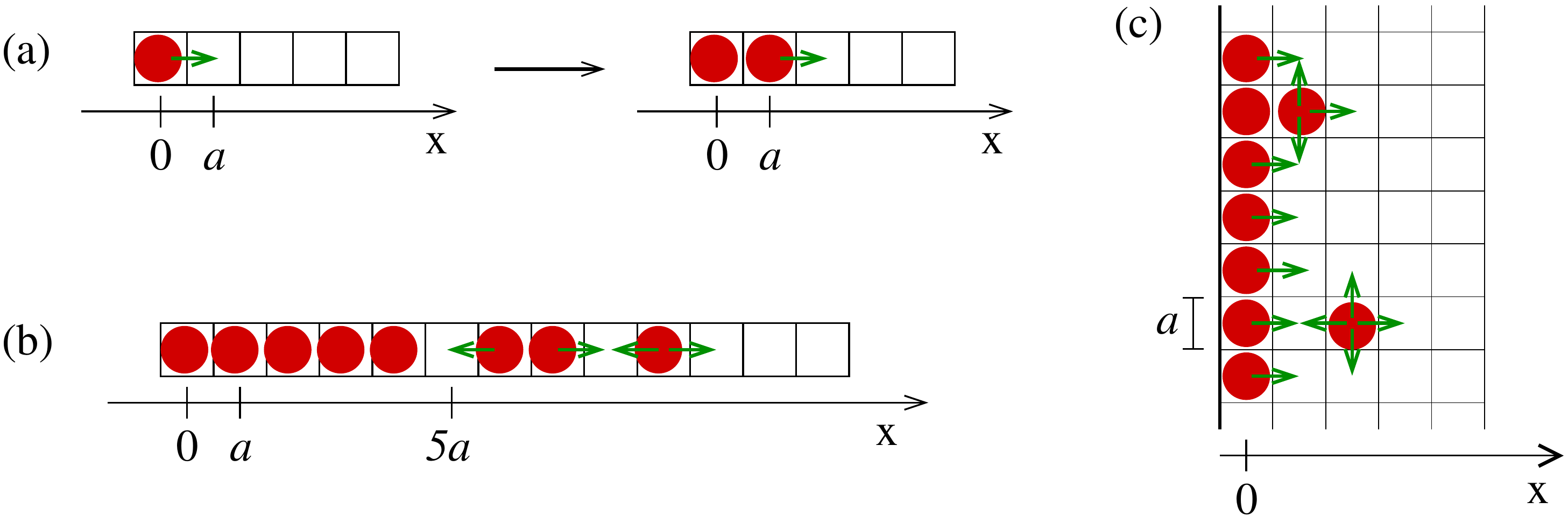}
\caption{
(a) Initial configurations of RWI in a line with a particle at the source $x=0$.
A new particle is inserted at the origin as soon as the initial particle moves
to the neighboring site.
(b) Configuration developed after some time, in which the particle at the source is blocked
and the closest vacancy is at $x=5a$.
(c) Configuration after a few steps of RWI in a square lattice with a line source at $x=0$.
The arrows indicate the possible hops of the infiltrated and of the source particles.
}
\label{diffmodel1d}
\end{figure}

Figs. \ref{diffmodel1d}(a),(b) illustrate the RWI in one dimension with a point
source at $x=0$ and all sites with $x>0$ empty at $t=0$.
As the particle at the source moves to $x=a$, another particle immediately refills the source
[Fig. \ref{diffmodel1d}(a)]; in this situation, only the particle at $x=a$ can move and the only
hop allowed for this particle is to the right.
Fig. \ref{diffmodel1d}(b) shows a configuration after several particles have entered the lattice,
in which only the particles at large distance from the origin can move.
The particle at the source is also unable to move in this situation; it will move only when the
site at $x=a$ is vacant, which requires diffusion of several intermediate particles.

An equivalent model is RWI in the region $x>0$ of a square lattice with a line source at $x=0$,
as shown in Fig. \ref{diffmodel1d}(c).
Again, when one particle leaves the line $x=0$, another particle is immediately added to that point.

For the simulation of RWI, we adopt an algorithm in which the particles that attempt to hop are
sequentially chosen at random from a list that includes all infiltrated particles and all particles
at the source.
Thus, there is no preferential choice of particles at the front or at the source for the first
hop attempts.
This brings to the model an additional randomization factor to that of the choice of the hop directions.

\subsection{Continuous limit}
\label{continuous}

Let $\rho\left( \vec{r},t\right)$ be the particle density at position $\vec{r}$ at time $t$,
given in number of particles per lattice site.
Here we derive an equation for its evolution in the continuous limit.

For simplicity, consider the one-dimensional case shown in Figs. \ref{diffmodel1d}(a),(b).
The density at position $x$ changes from time $t-\tau$ to time $t$ as the master equation
\begin{eqnarray}
\rho\left( x,t\right) - \rho\left( x,t-\tau\right) &=&
\frac{1}{2}\rho\left( x-a,t-\tau\right) \left[ 1 - \rho\left( x,t-\tau\right) \right] \nonumber \\
&& +\frac{1}{2}\rho\left( x+a,t-\tau\right) \left[ 1 - \rho\left( x,t-\tau\right) \right] \nonumber \\
&& -\frac{1}{2}\rho\left( x,t-\tau\right) \left[ 1 - \rho\left( x-a,t-\tau\right) \right] \nonumber \\
&& -\frac{1}{2}\rho\left( x,t-\tau\right) \left[ 1 - \rho\left( x+a,t-\tau\right) \right] .\nonumber \\
&&
\label{master}
\end{eqnarray}
In the right rand size of Eq. (\ref{master}), the first and second terms account for the
hops from neighboring sites to the position $x$, which are possible only if this site is not occupied;
the third and fourth terms account for the hops from site $x$ to neighboring sites, which are
possible only if those sites are not occupied.
The joint probabilities in Eq. (\ref{master}) were factorized, which considers that
the densities at neighboring positions are uncorrelated;
this assumption is similar to the independent cluster approximation which is also applicable for
problems of diffusion, aggregation, and fragmentation, with or without particle injection
\cite{majumdarPRL1998,reisstinchcombe2005a}.
This assumption cancels the crossed terms in the master equation.
In the continuous limit, we consider
$\tau \frac{\partial \rho}{\partial t}\approx \rho\left( x,t\right) - \rho\left( x,t-\tau\right)$
and $\frac{a^2}{2} \frac{\partial^2 \rho}{\partial x^2}\approx
\left[ \rho\left( x-a,t-\tau\right)-2\rho\left( x,t-\tau\right) + \rho\left( x+a,t-\tau\right)\right]$,
which give
\begin{equation}
\frac{\partial \rho}{\partial t} = D \frac{\partial^2 \rho}{\partial x^2} ,
\label{diffusioneq1d}
\end{equation}
where $D$ is given in Eq. (\ref{defD}).
This is the one-dimensional diffusion equation.
Generalization of the above arguments to higher dimensions is straightforward, which leads to
\begin{equation}
\frac{\partial \rho}{\partial t} = D\nabla^2 \rho .
\label{diffusioneq}
\end{equation}
The boundary conditions are $\rho =1$ at the source at all times
and $\rho\left( \vec{r},t=0\right) = 0$ at all points but the source.

Within the approximations in Eq. (\ref{master}), the excluded volume condition of the RWI model does
not affect the collective diffusion coefficient $D$ because the frustrated hop of a particle in
one direction is compensated by the frustration of the hop of another particle in the opposite direction.
However, if the tracer diffusion coefficient is measured (i. e. the coefficient characterizing the
motion of a particular particle), it is expected to be smaller than $D$ due to the frustration of
many hop attempts.

This continuous limit is expected to be valid only for $t\gg\tau$, where average particle
displacements are much larger than the lattice constant $a$.
In this case, the density $\rho\left( \vec{r},t\right)$ slowly varies in time
and we can compare RWI data with the solution of the diffusion equation, if available.

\subsection{Basic quantities}
\label{basic}

In RWI, we denote as $I\left( t\right)$ the number of infiltrated sites of the lattice at time $t$;
this quantity is hereafter termed the infiltration.
In the continuous description, $I$ is obtained by integration of the density $\rho$ through
the available space.
In one, two, and three-dimensions, $I$ respectively represent the total length, area, and volume
infiltrated by the random walkers.
The infiltration is expected to scale as a power law in time with exponent $n$, 
in a similar fashion the fluid infiltration in Eq. (\ref{defn})
(a discussion on the relation between RWI and fluid infiltration is postponed to Sec. \ref{fluid}).

We also calculate the root mean square (rms) distance of the infiltrated particles from the source,
$s_{RMS}\left( t\right)$; see illustration in Fig. \ref{infiltratedmedia}(a).
In RWI, a large fraction of the infiltrated particles move in regions with low density,
so that $s_{RMS}$ is expected to follow the same scaling law of free random walks in that medium:
\begin{equation}
s_{RMS} \sim t^{1/D_W} ,
\label{defDW}
\end{equation}
where $D_W$ is the random walk dimension.
In Euclidean lattices, we have normal diffusion with $D_W=2$, while in fractal lattices we
expect subdiffusion with $D_W>2$.

If the medium has a fractal dimension $D_F$ ($D_F=d$ in the cases of Euclidean lattices
in $d$ dimensions), then we define a characteristic infiltrated size $S$ by the equation
\begin{equation}
I = S^{D_F} .
\label{defS}
\end{equation}
$S$ may be interpreted as a characteristic radius of a region in which all infiltrated particles
are symmetrically located around the source with no empty site between them.
In fluid infiltration, the radius of the infiltrated region has the same scaling as $S$
(Sec. \ref{fluid}).
Figs. \ref{infiltratedmedia}(a),(b) illustrate this definition in effectively one-dimensional
infiltration problems.

The quantities calculated here are presented in terms of the dimensionless time
\begin{equation}
T\equiv \frac{t}{\tau} .
\label{defT}
\end{equation}
This is the average number of hop attempts of a particle that was located at the infiltration
boundary at $t=0$.

\subsection{Example: RWI in a square lattice with a line source}
\label{numerical1d}

\begin{figure}[!ht]
\center
\includegraphics[clip,width=0.45\textwidth,angle=0]{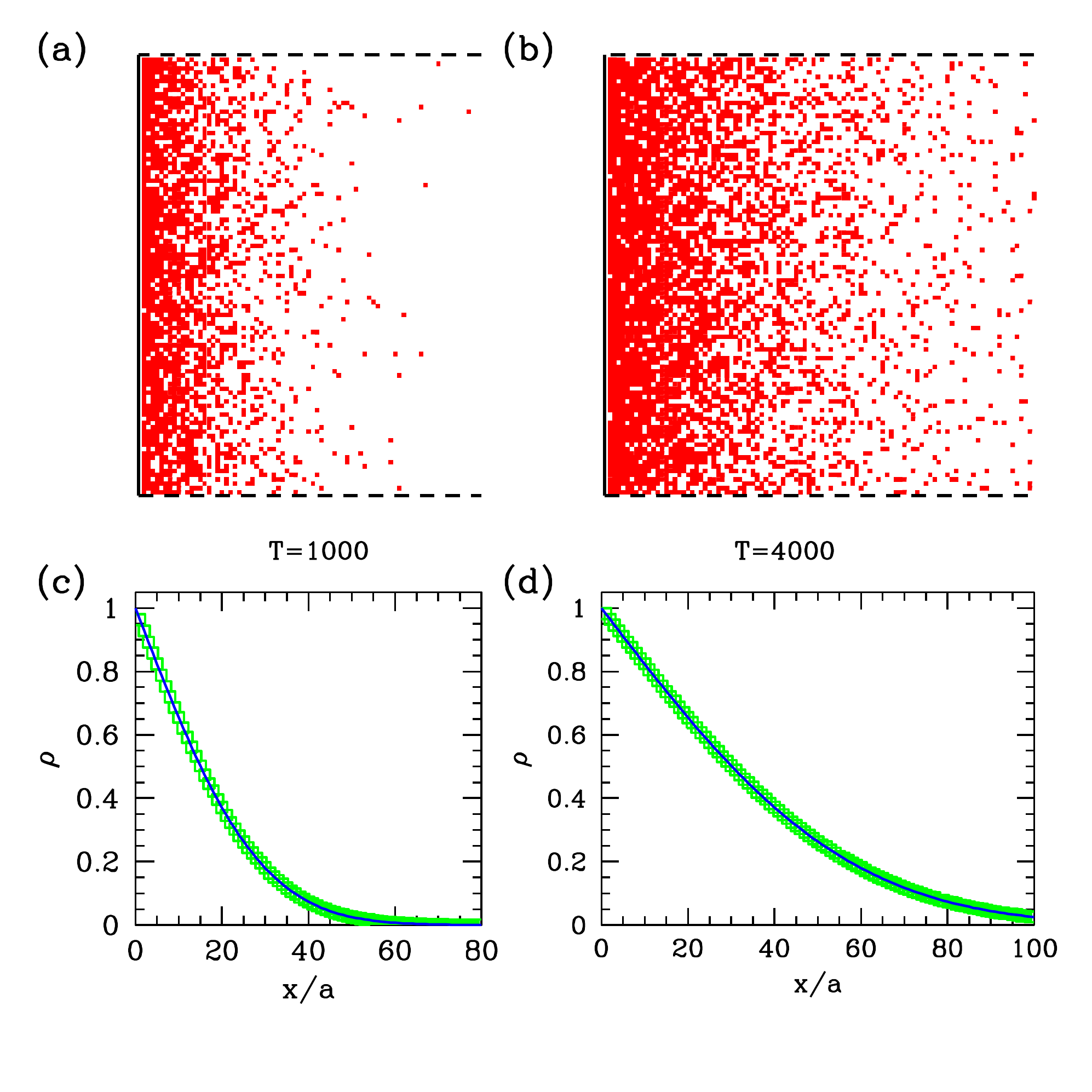}
\caption{
(a), (b) Sections of size $100a$ of an infiltrated square lattice with a line source at the left
in the two indicated times.
(c), (d) The corresponding concentration distributions calculated numerically (green squares) and 
analytically (blue curves).
}
\label{infsquare}
\end{figure}

Figs. \ref{infsquare}(a) and \ref{infsquare}(b) show configurations of an infiltrated square
lattice at $T=1000$ and $T=4000$, respectively, considering a line source at $x=0$.
Figs. \ref{infsquare}(c) and \ref{infsquare}(d) show the corresponding density distributions.
In each time, the numerical data are averages over $20$ configurations and source length $1000a$.
Figs. \ref{infsquare}(c) and \ref{infsquare}(d) also show the exact solution of
Eq. (\ref{diffusioneq1d}), $\rho\left( x,t\right) = 1-erf\left[ x/\left( 2\sqrt{Dt}\right)\right]$,
where $erf\left( u\right)$ denotes the error function of $u$.

There is a very good agreement between RWI data and the solution of the diffusion equation in this case.
This was formerly noted in Ref. \protect\cite{sapovalJPL1985}.
Discrepancies between the RWI and the continuous model are observed only at positions $x$ larger than
$\sqrt{Dt}$ by a factor of order $10$ [not shown in Figs. \ref{infsquare}(c),(d)],
since the diffusion equation fails to represent the RWI in these regions (e. g. the infinitely long
empty region in RWI due to the finite particle velocity).

The infiltration in this geometry scales as $I\sim T^{1/2}$ and Eq. (\ref{defS}) implies that the
characteristic infiltration length also scales as $S\sim T^{1/2}$.
This is the same scaling observed for the rms distance $s_{RMS}\left( t\right)$.

\subsection{Interpretation and relation with other models}
\label{relation}

RWI may represent the diffusive motion of a chemical species in a fluid that fills
the pores of a medium in contact with a reservoir (the source) of that species.
This interpretation suggests the application of the model to infiltration of solutes in rocks
or soils in cases where advective transport is negligible.
A particle in this model is not necessarily equivalent to a solute molecule, but may
represent a large number of molecules that entered the medium; in this case, the presence
or absence of a particle represents a local solute concentration above or below a given threshold,
respectively.
Consider, for instance, the application of this molecular interpretation to the configuration
of Fig. \ref{diffmodel1d}(b), in which particle hops from the source are not allowed due to
excluded volume effect.
It represents a case in which the region of the medium near the reservoir has a density
of molecules approximately equal to that of the reservoir.
In lengthscales smaller than the size of a lattice site, molecular diffusion has not ceased
inside and outside the medium, but the net flux from the source to the medium is zero.

The infiltrated region has many empty sites (vacancies) in the neighborhood of the infiltrated
particles, as illustrated in Figs. \ref{infsquare}(a) and \ref{infsquare}(b) for the case of a
line source in two dimensions.
The random walks of particles are equivalent to random walks of these vacancies, with excluded volume
conditions and with the same diffusion coefficient $D$.
When a vacancy reaches the source, it is immediately annihilated; this corresponds to the instantaneous
refilling of the source in RWI.
Thus, from the point of view of vacancy diffusion, we have a classical trapping problem
\cite{Havlin,rednerbook}.
Letting $\sigma\equiv 1-\rho$ be the density of vacancies, it also obeys the diffusion equation
(\ref{diffusioneq}); however, the boundary condition is $\sigma =0$ at the source.

The one-dimensional version of RWI [Fig. \ref{diffmodel1d}(a)] is equivalent to the trapping
of the vacancies by a static trap at the origin.
This model was originally proposed by Weiss, Kopelman, and Havlin \cite{weiss1989}.
The increase of the infiltration as $t^{1/2}$, which is observed in RWI,
is equivalent to the trapping rate decay as $t^{-1/2}$.
Despite the apparent simplicity, that trapping model has some nontrivial features; for instance,
the distance from the origin of the nearest vacancy increases anomalously as $t^{1/4}$
\cite{weiss1989}.
Similar anomalies were recently observed in other trapping models \cite{flekkoy2017}.

The square lattice version of RWI [Fig. \ref{diffmodel1d}(c)] was first proposed
by Sapoval, Rosso, and Gouyet \cite{sapovalJPL1985} for investigating the properties of the
interface of a conducting cluster formed by the particles connected to the source.
Subsequently, it motivated the proposal of the gradient percolation problem \cite{bunde1985,rosso1986}.
In a recent work, RWI in deterministic fractals was studied, also with the sources at flat boundaries
\cite{Reis2016}, which helped to understand scaling properties of infiltration models.

\section{RWI with localized sources}
\label{localized}

\begin{figure}[!ht]
\center
\includegraphics[clip,width=.45\textwidth,angle=0]{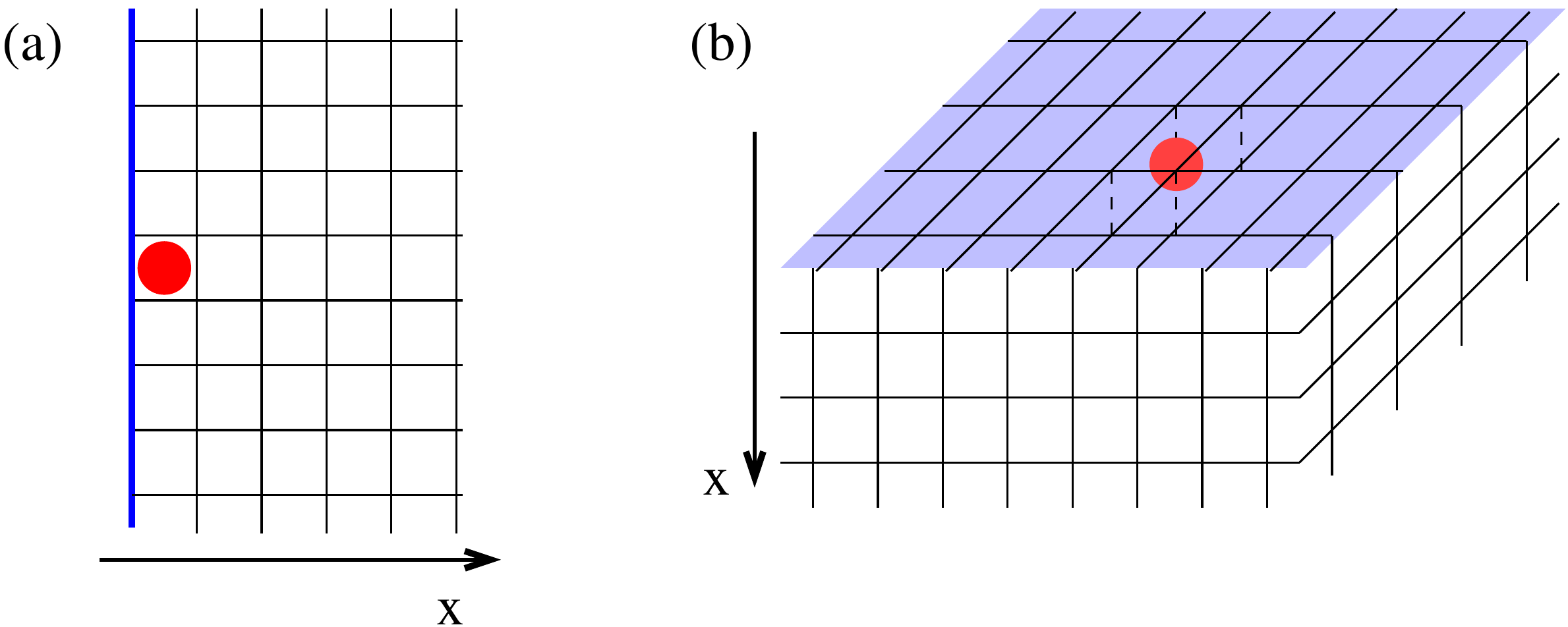}
\caption{
Initial condition of RWI with a point source (red circle and sphere) and an impenetrable
boundary at $x=0$ (blue) in (a) two and (b) three dimensions.
}
\label{diffmodel2d3d}
\end{figure}

Here we consider the RWI models defined in Sec. \ref{definition}
with a point source localized in the neighborhood of an impenetrable boundary,
as illustrated in Figs. \ref{diffmodel2d3d}(a) ($d=2$) and \ref{diffmodel2d3d}(b) ($d=3$).
This geometry is chosen to represent a large porous medium with a narrow hole at a
surface which is in contact with a reservoir of the solute.
The same geometry is also considered in simulations of RWI in determininistic fractals
embedded in two- and three-dimensional lattices.

For simplicity, solutions of the diffusion equation (\ref{diffusioneq}) will be obtained
with the source in media that are infinitely large in all directions (i.e., without the
impenetrable boundary).
This is expected to provide the same time scaling of the relevant quantities.

\subsection{Infiltration in homogeneous media}
\label{homogeneous}

\subsubsection{Two dimensions}
\label{twodimensions}

In this case, an equivalent diffusion model considers an infinite line source of radius $a$
and constant density $\rho(r \le a, t\geq 0)=1$ placed in an infinite three-dimensional domain
initially with $\rho(r>a,t=0)=0$.
The subsequent change in the density in a plain of fixed $z$ is governed by the diffusion equation
\begin{equation}
\label{eqdiff2d}
\frac{\partial \rho}{\partial t}=\frac{D}{r} \frac{\partial}{\partial r}\left(r\frac{\partial \rho}
{\partial r} \right), \ \ \ r>a, \ \rho\left(r \rightarrow \infty, t\right)=0 .
\end{equation}    

Ref. \protect\cite{crank} (pp. 87-88) provides a closed solution for this problem,
from which we can calculate the amount fluxed into the region $r>a$ per unit area and time,
$F\equiv -D\left(\frac{\partial \rho}{\partial r} \right)_{r=a}$.
Assuming for convenience that $D=1/4$ and $a=1$ [which leads to $T=t$ in Eq.(\ref{defT})],
the solution for large times is
\begin{equation}
\label{flux2d}
F=\frac{1}{2} \left[ \left(\frac{1}{\ln{T}-2\gamma}  -\frac{1}{(\ln{T}-2\gamma)^2} \right)
+\frac{1-\gamma} {(\ln{T}-2\gamma)^2}  \ldots \right] ,
\end{equation}  
where $\gamma=0.57722$ is Euler's constant and $T$ is defined in Eq. (\ref{defT}).
From this we can calculate the infiltration through a simple integration in time;
this is denoted as $I_{DE}$ to stress that it results from the solution of a diffusion equation.
Integrating the first bracketed term on the right hand side of Eq. (\ref{flux2d}),
the infiltration in the half plane $x\geq 0$ [as in the RWI geometry of Fig. \ref{diffmodel2d3d}(a)]
is, to leading order,
\begin{equation}
\label{I2d}
I_{DE}\left( t\right) = \pi \int_0^t F dT = \frac{\pi}{2} \frac{T}{\log(T)-2\gamma} .
\end{equation}    

We performed simulations of RWI in square lattices of lateral size $L=2000$ with a point source in
the middle of a border up to $T=2\times {10}^4$, which led to the injection of more than
$3\times {10}^3$ particles.
The number of configurations used to calculate $I$ and $s_{RMS}$ was ${10}^3$ (i. e. this is the number of
times in which particles were injected in an empty lattice).

\begin{figure}[!ht]
\center
\includegraphics[clip,width=0.45\textwidth,angle=0]{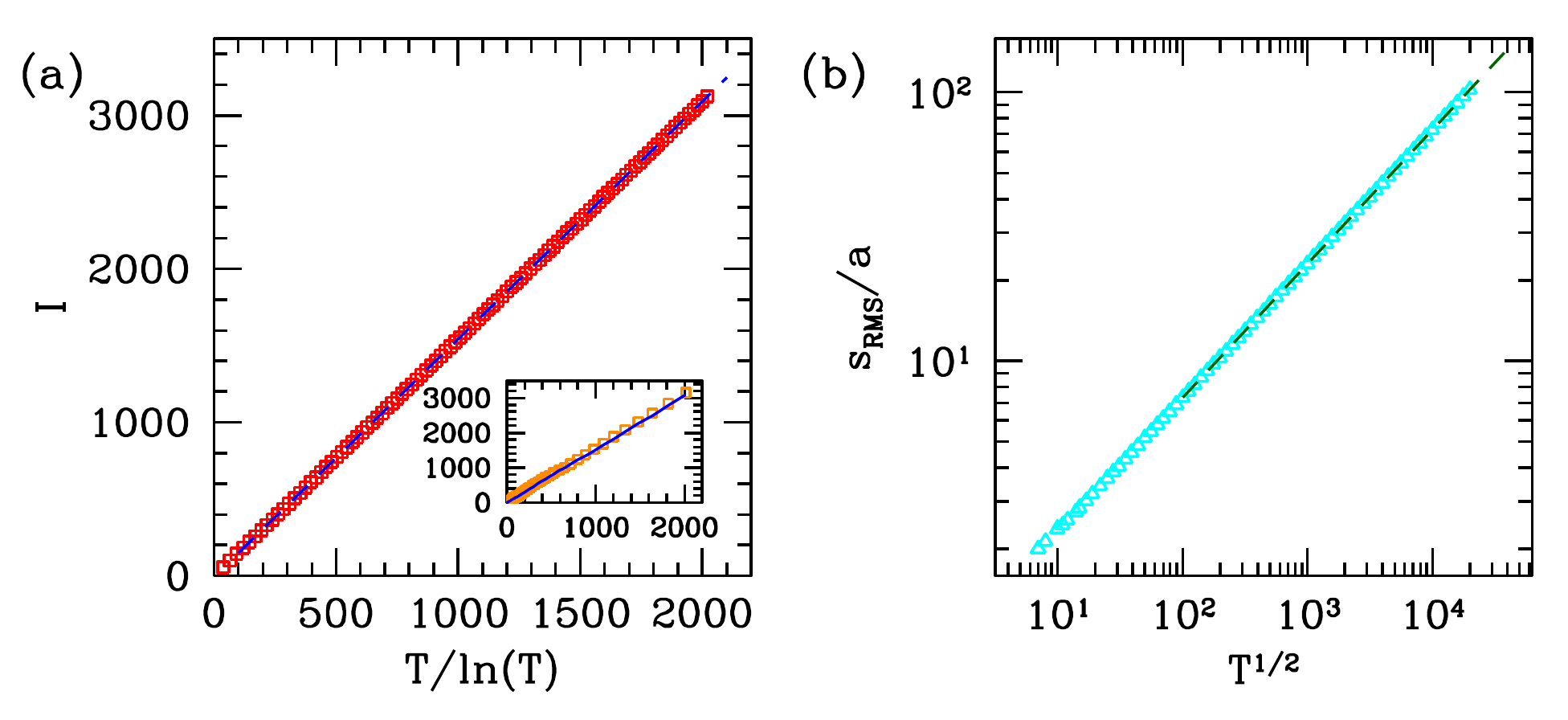}
\caption{
RWI from a point source in two dimensions:
(a) infiltration versus $T/\ln{T}$ (red squares) and a linear fit of the data (dashed blue line);
(b) rms distance from the source versus $T^{1/2}$ (blue triangles) and a linear fit of the data
(dashed green line) whose slope is $0.499$.
The inset in (a) compares results in two configurations (blue line and orange squares).
}
\label{infilt2d}
\end{figure}

Fig. \ref{infilt2d}(a) shows $I$ (the number of infiltrated lattice sites) as a function of $T/\ln{T}$.
The inset of Fig. \ref{infilt2d}(a) shows that the data obtained in two different configurations are
very close to each other.
This means that the accuracy of the averaged data is very high, so that the standard deviations in
the main plot are much smaller than the size of the data points.
The excellent linear fit in that plot confirms that $I$ has the same 
logarithmic corrected time scaling predicted by the continuous model [Eq. (\ref{I2d})].
Fig. \ref{infilt2d}(b) shows a bilogarithmic plot of $s_{RMS}$ as a function of $T$, which
confirms the diffusive increase of that average displacement; the accuracy was similar to that of $I$.

The characteristic length of the infiltration [Eq. (\ref{defS})] scales as
$S\sim {\left( T/\ln{T}\right)}^{1/2}$.
Thus, it is smaller than the rms distance $s_{RMS}$ due to the logarithmic correction.

\subsubsection{Three dimensions}
\label{threedimensions}

In this case, the equivalent diffusion model has a spherical point source of radius $a$ and 
constant density $\rho\left(r \le a, t>0\right)=1$, placed in a infinite domain initially
with $\rho\left( r>a,t=0\right) =0$.
The subsequent change in the potential in $r>a$ is governed by the diffusion equation
\begin{equation}
\label{diffeq3d}
\frac{\partial \rho}{\partial t}=\frac{D}{r^2} \frac{\partial}{\partial r}
\left(r^2\frac{\partial \rho}{\partial r} \right) , \ \ \ r>a, \
u\left( r \rightarrow \infty, t\right) =0
\end{equation}    
The analytical solution (Ref. \protect\cite{crank}, pp. 102-103) is    
\begin{equation}
\label{rho3d}
\rho=\frac{a}{r} \text{erfc}\left(\frac{r-a}{2\sqrt{Dt}}\right)
\end{equation}    
The amount fluxed into the region $r>a$ per unit area and time is given as
\begin{equation}\label{F3d}
F=-D\left(\frac{\partial u}{\partial r} \right)_{r=a}=\frac{D}{a} + \sqrt{\frac{D}{\pi t}}
\end{equation}    
From this we can calculate the infiltration $I_{DE}$ through a simple integration in time.
Considering $D=1/6$ and $a=1$ [which leads to $T=t$ in Eq.(\ref{defT})] and
integrating only in the region $x\geq 0$ [as in Fig. \ref{diffmodel2d3d}(b)], we obtain
\begin{equation}\label{I3d}
I_{DE} = 2 \pi \int_0^t F dt = \frac{\pi}{3} T+\frac{4 \sqrt{\pi}}{ \sqrt{T}}  .
\end{equation}    

Our simulations of RWI were run up to $T=2\times {10}^4$ in a lattice with lateral size $L=700$;
$I$ and $s_{RMS}$ were averaged over ${10}^3$ configurations, which also provides
these quantities with high accuracy.
Approximately ${10}^4$ particles were injected in each configuration.

\begin{figure}[!ht]
\center
\includegraphics[clip,width=0.45\textwidth,angle=0]{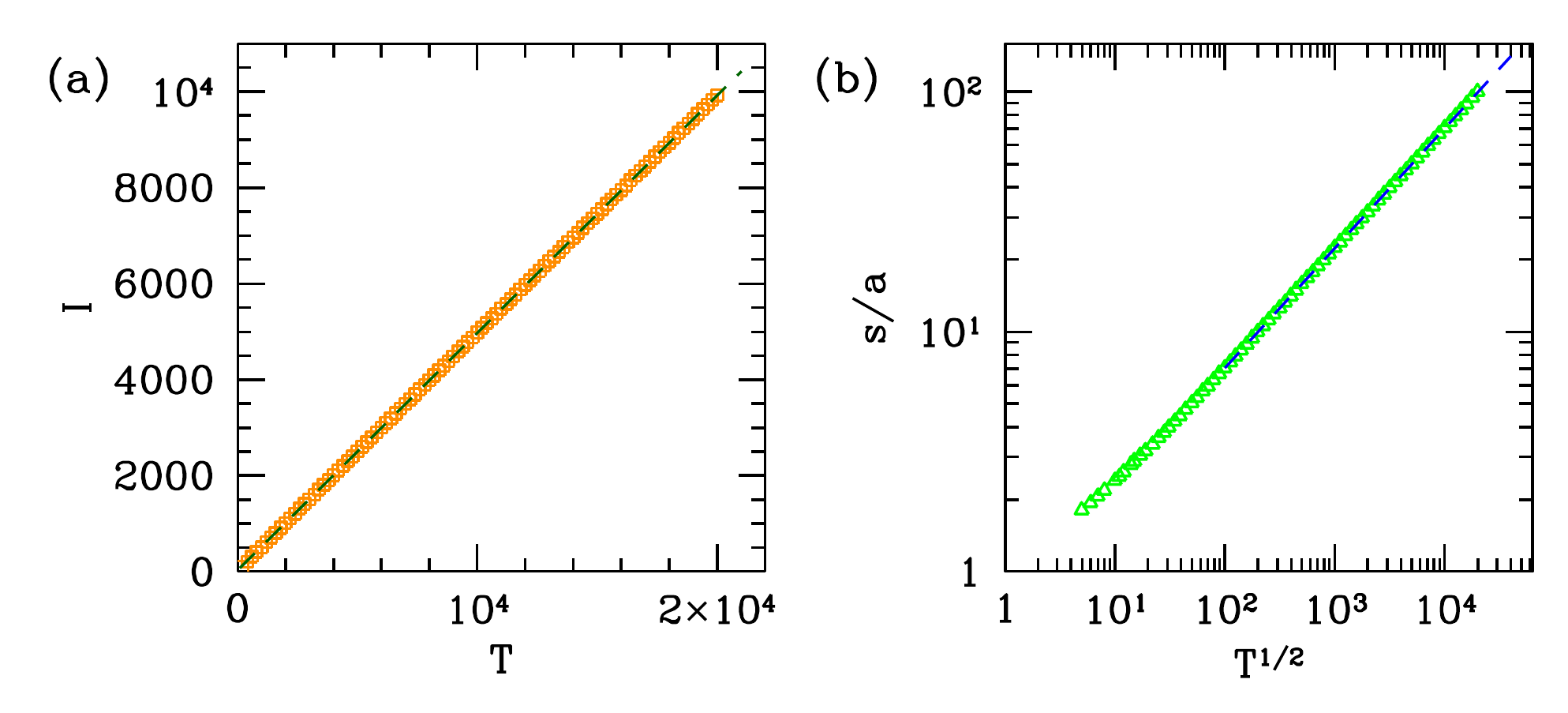}
\caption{
Results for infiltration from a point source in three dimensions:
(a) infiltration versus $T$ (orange squares) and a linear fit of the data (dashed green line);
(b) rms distance from the source versus $T$ (green triangles) and a linear fit of the data
(dashed blue line) whose slope is $0.500$.
}
\label{infilt3d}
\end{figure}

Fig. \ref{infilt3d}(a) shows $I$ as a function of $T$; the linear fit confirms that it has
the same scaling as $I_{DE}$ in Eq. (\ref{I3d}).
The characteristic length of the infiltration [Eq. (\ref{defS})] consequently scales as
$S\sim T^{1/3}$.
Fig. \ref{infilt3d}(b) shows a bilogarithmic plot of $s_{RMS}$ as a function of $T$, which
confirms the diffusive increase of that average distance.
In this case, we also observe that the characteristic length $S$ defined from the infiltrated
volume does not have the diffusive scaling of the rms distance from the source.

\subsection{Scaling approach for the infiltration}
\label{scaling}

Here we explore the observation of Sec. \ref{relation} that RWI is equivalent to the problem
of trapping of vacancies at the source.
It implies that the infiltration rate $\frac{\text{d}I}{\text{d}t}$ is equal to the probability
$P_{trapping}\left( t\right)$ of a vacancy to be trapped at the source at time $t$.
This is the probability that the first passage of the vacancy at the source occurs at time $t$.
For long times, it is independent of the source position,
but it is strongly dependent on the spatial dimension \cite{montroll,rednerbook}.

$I$ is obtained by the integration of the first passage probability.
Since this probability can be obtained as a property of a single random walk, $I$ is equal to
the sum of this probability over the hops of a single walker up to time $t$.
When the walker hops to a site for the first time, the number of distinct visited sites increases
by one unit; thus, $I\left( t\right)$ equals the number of distinct sites visited by the walker
in the medium up to time $t$, which is denoted as $N_D\left( t\right)$ \cite{montroll,rednerbook}.
The infiltration exponent $n$ in Eq. (\ref{defn}) is the same exponent of the time scaling
of the number of distinct sites visited by a random walker; an amplitude relating $I$ and
$t^n$, which is omitted in Eq. (\ref{defn}), may depend on the boundary conditions.

In two dimensions, the dominant term in the scaling of the number of distinct sites visited
by a random walker is $N_D\sim t/\ln{t}$ \cite{montroll}; this agrees with the simulation
result for the RWI in Sec. \ref{twodimensions}.
In three dimensions, the dominant term is $N_D\sim t$ \cite{montroll}, which also agrees with
our simulation results for RWI (Sec. \ref{threedimensions}).

These results establish a relation between RWI and recurrence properties of lattice random walks.
Moreover, since the time evolution of the infiltration in RWI may also be predicted by the
diffusion equation with constant concentration at the source, our approach also connects
the solution of this equation with the recurrence properties of random walks.

\subsection{Infiltration in Sierpinski carpets}
\label{carpets}

Here we study RWI infiltration with localized sources in fractals embedded in two
dimensions combining the previous scaling approach, numerical results of RWI, and results
of numerical integration of the diffusion equation.

\begin{figure}[!ht]
\center
\includegraphics[clip,width=.45\textwidth,angle=0]{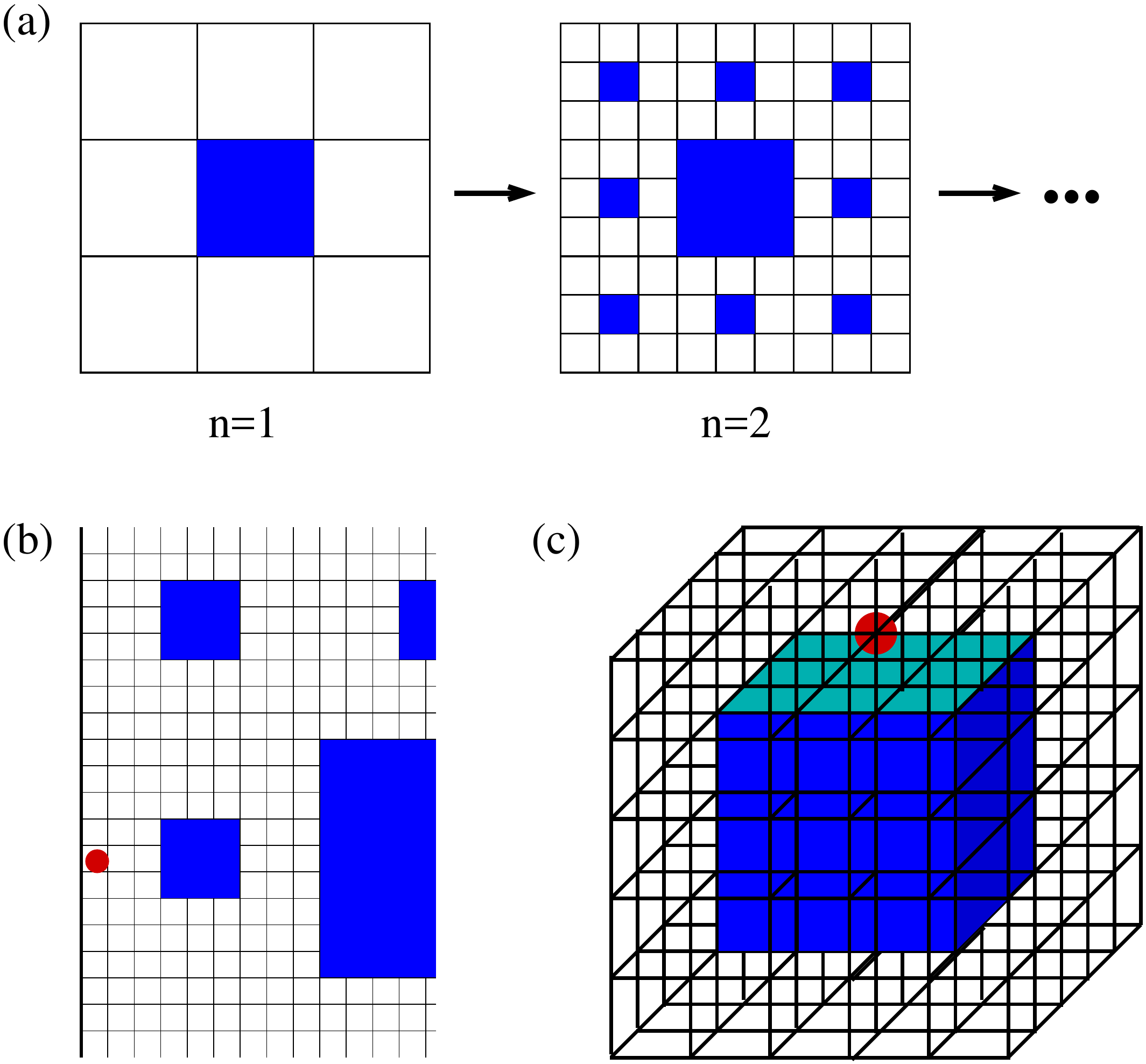}
\caption{
(a) First two stages of construction of the carpet with $b=3$ and $m=1$, with the impenetrable
solid in blue.
(b) Initial condition for RWI in a carpet with a point source (red) at an outer boundary.
(c) Initial condition for RWI in a sponge with a point source (red) at an outer boundary.
}
\label{diffmodelcarpet}
\end{figure}

The iterative construction of a Sierpinski carpet is illustrated in Fig. \ref{diffmodelcarpet}(a).
We denote as $b$ the scaling factor of the carpet and $m$ the number of 
subsquares removed in the first iteration of the construction.
At each stage of construction, each subsquare is replaced by the generator (stage $n=1$).
The removed subsquares represent the impenetrable (solid) part of the medium and the empty
subsquares form the porous medium.
The length of an outer border of the carpet after $n$ iterations is $L=b^n$.
The fractal is obtained after an infinite number of iterations, with fractal dimension
$D_F=\ln{\left( b^2-m\right)}/\ln{b}$.
The initial condition for RWI in a carpet with a point source is shown in
Fig. \ref{diffmodelcarpet}(b).

The number of distinct sites visited by a random walker in this fractal is expected to scale as
\begin{equation}
N_D \sim t^{D_S/2} \qquad , \qquad D_S = 2 \frac{D_F}{D_W} ,
\label{AOrelation}
\end{equation}
where $D_S$ is the spectral (or fracton) dimension.
Eq. (\ref{AOrelation}) is known as Alexander-Orbach (AO) relation \cite{AO,Havlin}; it is based on
the assumption of homogeneous distribution of the random walker position within the accessible
fractal volume at time $t$.
However, the AO relation has been criticized in the past due to deviations observed in some
numerical results; for a discussion, see Ref. \protect\cite{Havlin}, Ch. 7,
and Ref. \protect\cite{sokolov}.

For RWI, Eqs. (\ref{defn}) and (\ref{AOrelation}) give the infiltration exponent $n=D_F/D_W$.
In the carpet of Fig. \ref{diffmodelcarpet}(a), $D_F=\ln{8}/\ln{3}=1.8928\dots$.
Ref. \protect\cite{balankin2017} conjectured that the exact value of the random walk dimension
in this carpet is $D_W=2+\ln{\left( 10/9\right)}/\ln{3} =2.0959\dots$; this value agrees with the
numerical estimates $2.101\left( 22\right)$ \cite{reisJPA1995} and
$2.106\left( 16\right)$ \cite{kimJPA1993}.
Using the exact $D_F$ and the conjectured $D_W$, we obtain $n=0.903\dots$.

We performed simulations of RWI in the $7^{\text{th}}$ stage of construction of this carpet 
up to $T={10}^5$.
Averages were also taken over ${10}^3$ configurations, with more than ${10}^4$
injected particles in each one; thus, the accuracy of the data is similar to that in
Fig. \ref{infilt2d}.

\begin{figure}[!ht]
\center
\includegraphics[clip,width=.45\textwidth,angle=0]{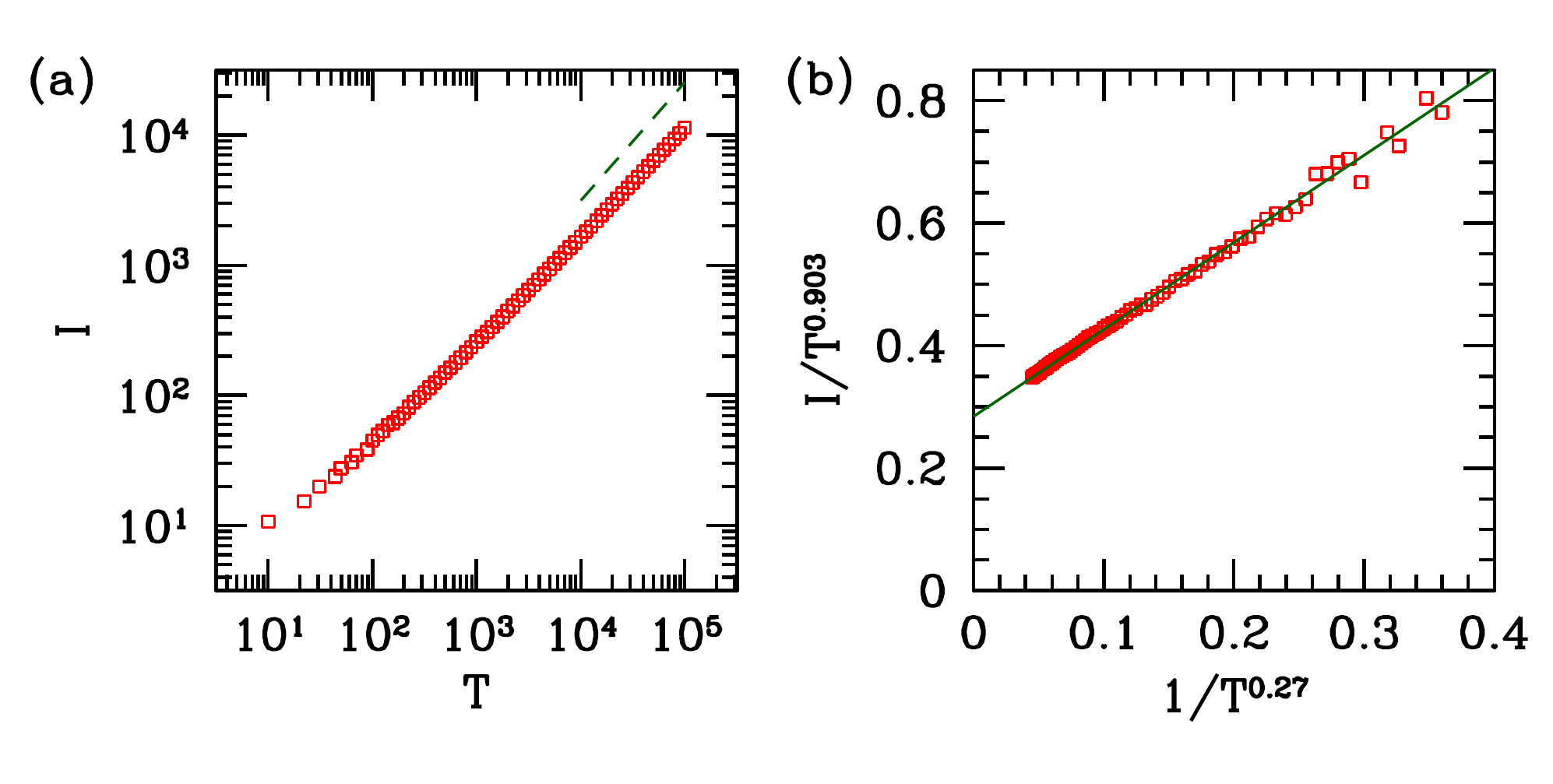}
\caption{
(a) Time evolution of the infiltration from a point source in the SC with $b=3$ and $m=1$.
The dashed line has the theoretically predicted slope $0.903$.
(b) Scaled infiltration as a function of $1/t^{0.27}$ for RWI in the Sierpinski carpet.
The solid line is a linear fit of the data for $5\times {10}^2\leq T\leq {10}^5$.
}
\label{infpont31}
\end{figure}

In Fig. \ref{infpont31}(a), we show a bilogarithmic plot of $I$ versus $T$ in this fractal.
The slope at long times is still slightly smaller than the proposed value of $n$.
This is consistent with previous calculations of $N_D$ in the carpets, which suggested corrections
to the AO relation \cite{reisPLA1996,dasgupta1999}.
In Fig. \ref{infpont31}(b), we show the ratio $I/T^{0.903}$  as a function of $1/T^{0.27}$;
the variable in the abscissa is the one that provides the best linear fit of the data in the
range $5\times {10}^2\leq T\leq {10}^5$ (among other variables in the form $T^{-\lambda}$ with
positive $\lambda$).
The convergence to a finite value at long times confirms that the infiltration scales
asymptotically with the proposed value of $n$, and the small exponent in the abscissa of
Fig. \ref{infpont31}(b) indicates the presence of large subdominant corrections to the AO relation.

For comparing different methods of solution for the RWI problem,
we also solved Eq. (\ref{diffusioneq}) numerically.
We use a time explicit finite-difference control volume solution for the
solute infiltration from a point source ($\rho =1$) into the stage $n=3$ of the same
Sierpinski carpet [Fig. \ref{diffmodelcarpet}(a)], with diffsuivity $D=1$ at the empty subsquares
and with the solid squares representing areas of very low diffusivity $D<<1$.
The numerical solutions, similar to that outlined in  \cite{Voller_wrr}, involves a $3^6 \times 3^6$
mesh of square node-centered control volumes of side length $\Delta =1/3^6$ and, to retain stability
of the explicit scheme, a  time step $\Delta t = 0.125\times\Delta^2$.  

\begin{figure}[!ht]
\center
\includegraphics[clip,width=.5\textwidth,angle=0]{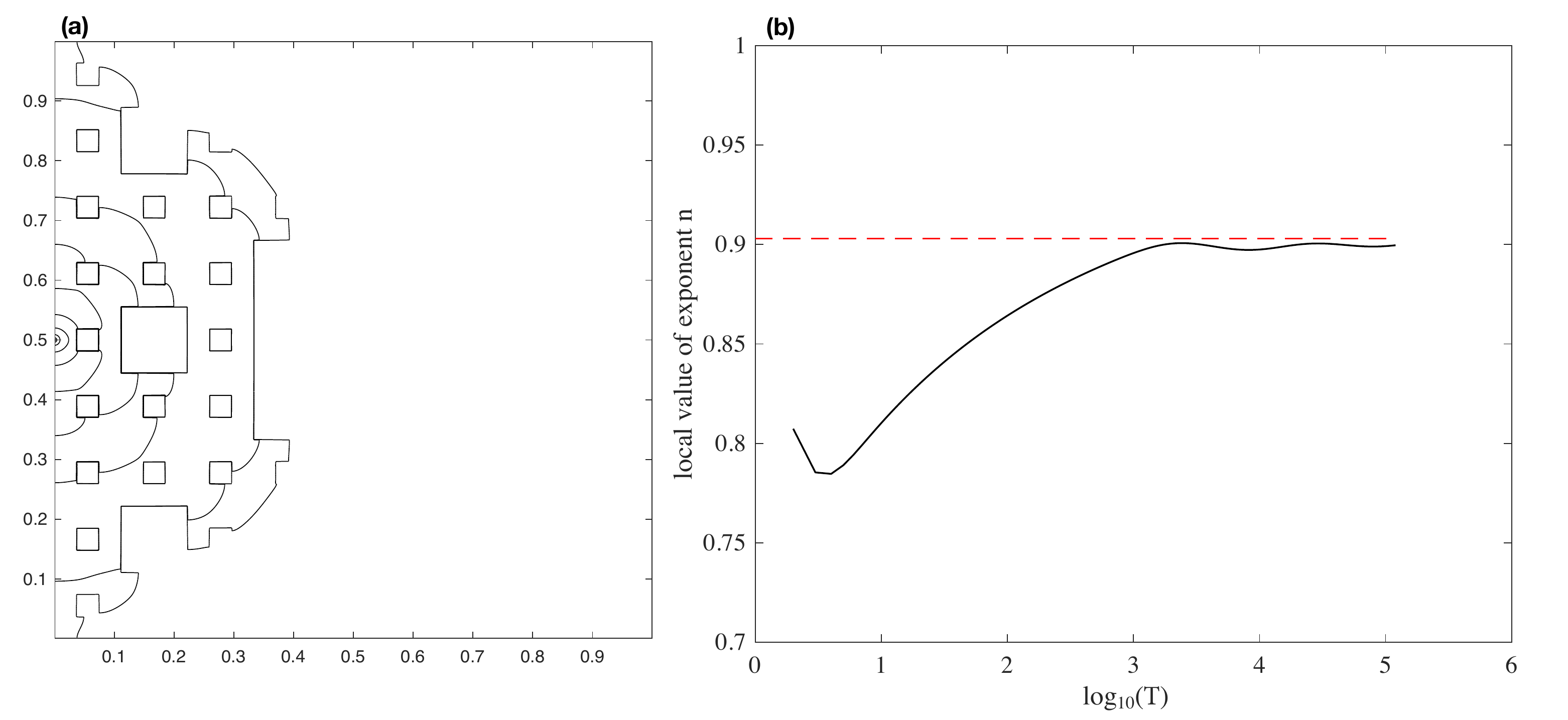}
\caption{
Solute diffusion into a 3$^{\mbox{rd}}$ stage Sierpinski carpet with $b=3, m=1$.
(a) The concentration contours after 120,000 time steps: from right to left,
$\rho=0.0.005$, $0.01$, $0.05$, and from $0.1$ to $0.9$ in intervals of $0.1$.
Solid squares and pores are not distinguished by different colors here.
(b) The local time exponent $n= d{\text ln}I_{DE}/d{\text ln}t$; the dashed line has the theoretically
predicted value $0.903$.}
\label{VVadd1REV}
\end{figure}

Fig. \ref{VVadd1REV}(a) shows some concentration contours after 120,000 time steps.
The measure over time of this simulation is the solute content $I_{DE}$, see Eq. (\ref{I2d}),
from which we obtain the local time exponent $n= d{\text ln}{I_{DE}}/d{\text ln}t$.
The key observation in Fig. \ref{VVadd1REV}(b) is that this exponent approaches,
in long times, a consatnt  value that is within fractions of a percent of the expected value
of $\sim$0.903.
Consequently, this solution significantly improves that of direct simulation of RWI
(Fig. \ref{infpont31}).
More importantly, it confirms with very good accuracy
that the AO relation is valid in a Sierpinski carpet,
advancing over the current and previous random walk based methods.
      
The oscillations observed in Fig. \ref{VVadd1REV}(b) at long times resemble the
log-periodic oscillations shown in Refs. \protect\cite{babJCP2008,akkermans} in models
of random walks in regular fractals.
Some methods were developed to treat data with this type of oscillation \cite{babJCP2008,haber2014},
which may be helpful whenever accurate exponent estimates are not available.

In Ref. \protect\cite{Reis2016}, a scaling approach was used to relate the infiltration exponent
$n$ [Eq. (\ref{defn})] and the random walk dimension $D_W$ in cases where the source was
one of the outer boundaries of the fractal.
For a border with dimension $D_B$, it was shown that
\begin{equation}
n = \frac{D_F -D_B}{D_W} .
\label{scalingnnu}
\end{equation}
This result may be extended to infiltration from a point source in a carpet because, in this case,
$D_B=0$, which gives $n=D_F/D_W$.

\subsection{RWI in Menger sponges}
\label{sponge}

A Menger sponge is iteratively constructed here by extending to three dimensions the
procedure used to construct the SCs:
first, a cube is divided in $b^3$ subcubes and $m$ symmetrically distributed subcubes are removed,
which forms the generator; each one of the subcubes is then divided following the same rule,
and this process is repeated indefinitely.
The fractal dimension of the sponge is $D_F=\ln{\left( b^3-m\right)}/\ln{b}$.

Here we study RWI in the sponge with $b=5$ and $m=27$.
The initial condition for RWI with a point source in this sponge
is illustrated in Fig. \ref{diffmodelcarpet}(c).
The fractal dimension is $D_F=\ln{98}/\ln{5}\approx 2.84880$ and the
estimate of the random walk dimension is $D_W = 2.09(6)$ \cite{Reis2016}.

These values imply $D_F > D_W$.
The AO relation [Eq. (\ref{AOrelation})] is expected to be valid only when random walks are
recurrent in a given fractal, but this is not the case here.
Instead, here the accessible region for the infiltrated particles has a radius $\sim s_{RMS}$ and
a volume of order $s_{RMS}^{D_F}\sim t^{D_F/D_W}\gg t$, i.e. this region is much larger than the
number of sites that the walker can visit within a time $t$.
This means that only a small fraction of the accessible sites is visited, and this fraction tends
to zero asymptotically.
In such cases, we expect that $N_D$ scales linearly in time, and so $I$ is also linear in time.

\begin{figure}[!ht]
\center
\includegraphics[clip,width=.35\textwidth,angle=0]{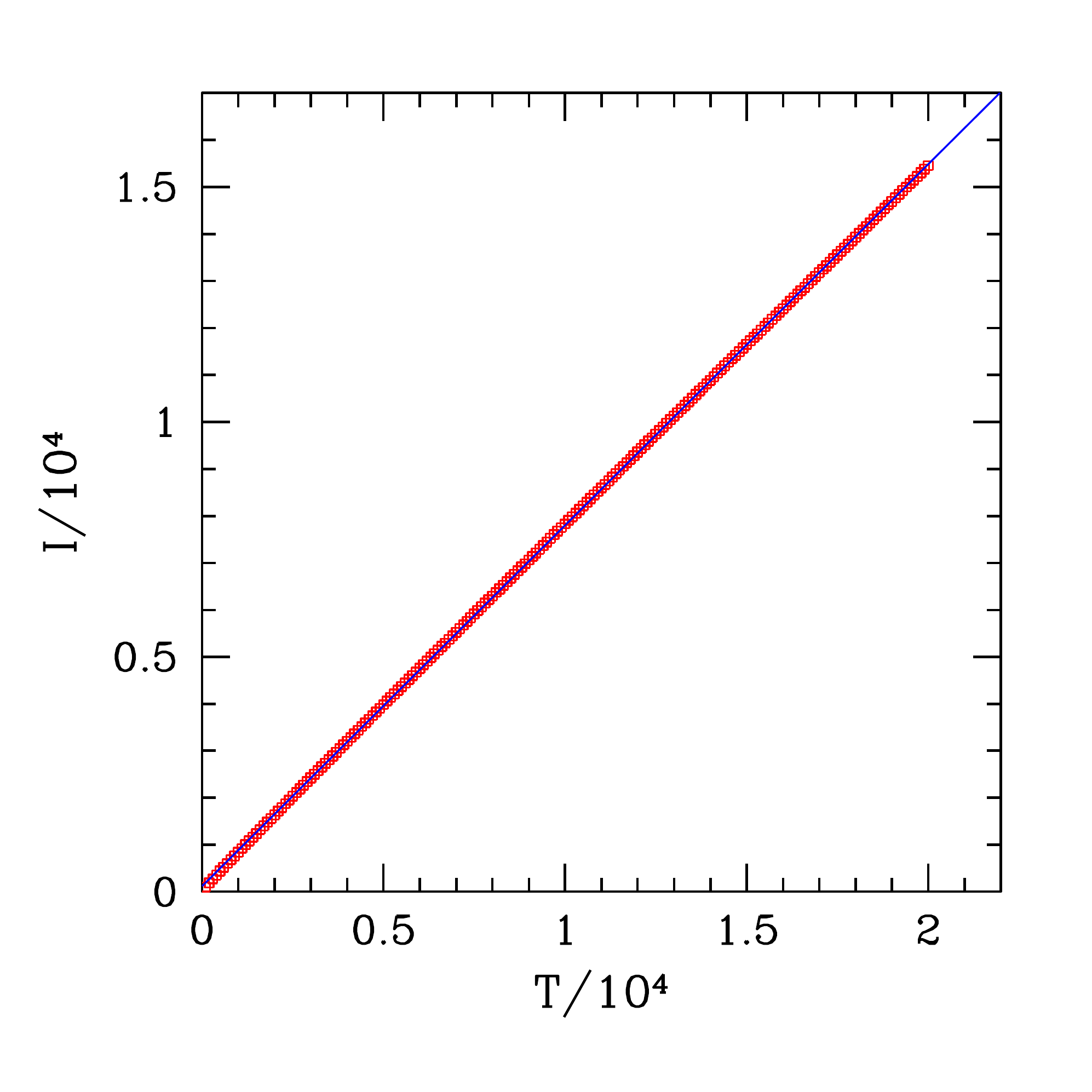}
\caption{Infiltrated volume as a function of time for RWI in the Menger sponge.
The solid line is a linear fit of the data.
}
\label{infpontm53}
\end{figure}

We simulated RWI in the $4^{\text{th}}$ stage of construction of the sponge up to $T=2\times {10}^4$.
Averages were taken over ${10}^3$ configurations and more than $1.5\times {10}^4$ particles were
injected in each configuration.
Fig. \ref{infpontm53} shows that the infiltrated volume $I$ increases linearly in time in all
the simulated time interval, which confirms the above prediction.
Note that the same scaling was obtained in a homogeneous three-dimensional medium with a point source,
as shown in Sec. \ref{threedimensions}.

It is interesting to observe that the infiltration in this fractal is normal,
although the diffusion of a single particle is clearly anomalous ($D_W >2$).
The same observation is applicable to other fractals where random walks are not recurrent,
i.e. $D_W<D_F$.
Consequently, in an infiltration experiment on such a fractal with a localized source, the
measurements of infiltrated mass or volume will not show any anomaly, but the same linear increase
observed in a homogeneous medium.
In order to observe the anomaly, it is necessary to measure the diffusion lengths
of the infiltrated molecules (or of other tracer particles).

RWI in the sponge also shows a failure of the scaling relation (\ref{scalingnnu}), since that relation
implies the AO relation for $D_B=0$.
Indeed, Eq. (\ref{scalingnnu}) was obtained in Ref. \protect\cite{Reis2016} under the assumption
that a finite fraction of a region of lateral size $s_{RMS}$ was filled with infiltrated particles.
This is actually the case in two-dimensional systems with linear sources or three-dimensional
systems with planar sources.
However, in a fractal with large $D_F$, a point source is too small to fill a finite fraction
of a region with that size.

\section{Relation with fluid infiltration}
\label{fluid}

Here we consider infiltration of a fluid with a constant pressure head $h=1$ at a fixed boundary,
which is the source of the fluid molecules, and assume that the head vanishes at the infiltration
front.
In the region limited by these boundaries, the slowly varying pressure head $h$ obeys the
Laplace equation
\begin{equation}
\nabla^2 h = 0 ,
\label{laplacianh}
\end{equation}
and the velocity of the infiltration front obeys the Darcy law
\begin{equation}
\vec{v} = -K \vec{\nabla} h , 
\label{darcy}
\end{equation}
where $K$ is the hydraulic conductivity.
For simplicity, we consider $K=1$ in a homogeneous medium.

This model was numerically studied in several carpets with different boundaries in Refs.
\protect\cite{Voller_wrr} and \protect\cite{reisbolstervoller}.
The estimates of the infiltration exponent $n$ were in excellent agreement with estimates from
the RWI model and with the scaling relation (\ref{scalingnnu}) \cite{Reis2016}.
Experiments with infiltration in finite stages of construction of some carpets
also showed good agreement with those estimates \cite{Voller_wrr2}.

Here we extend this comparison to infiltration with localized sources.

\subsection{Two dimensions and localized source}
\label{fluid2d}

We consider the problem in a two-dimensional cylindrical geometry, where a small source is
located at $r=a$.
Eq. (\ref{laplacianh}) becomes
\begin{equation}
\frac{\partial}{\partial r}\left(r \frac{\partial h}{\partial r}\right)=0, \  \  \ h(a)=1, \ h(R)=0 ,
\label{eqh2dpoint}
\end{equation}
which gives the head profile
\begin{equation}
h=\frac{\ln{r}- \ln{R}}{\ln{a} -\ln{R}}, \  \  \ a\le r \le R .
\label{h2dpoint}
\end{equation}
Eq. (\ref{darcy}) for the moving front becomes
\begin{equation}
- \left(\frac{\partial h}{\partial r} \right)_{r=R}=\frac{dR}{dt} 
\label{front2dpoint}
\end{equation}
and substitution of Eq. (\ref{h2dpoint}) gives
\begin{equation}
\frac{dR}{dt}=-\frac{1}{R\left(\ln{a} - \ln{R}\right) }, \  \  \ R\left( 0\right) =a
\label{eqR2dpoint}
\end{equation}
The solution for $R(t)$ is
\begin{equation}
\frac{R^2}{2} \left(\ln{R}- \log a \right)+\frac{a^2-R^2}{4}=t
\label{R2dpoint}
\end{equation}
In limit of $R>>a$, i.e. when the front displacement is much larger than the radius of the source,
we have
\begin{equation}
R\approx 2{\left( \frac{t}{\ln{t}} \right)}^{\frac{1}{2}}  .
\label{R2dpointapprox}
\end{equation}
The infiltrated area in this limit is
\begin{equation}
A\approx 4\pi\left( \frac{t}{\ln{t}} \right)  .
\label{A2dpointapprox}
\end{equation}

This is the same linear increase depleted by a logarithmic correction that we observed in RWI
(Sec. \ref{twodimensions}).
The front radius $R$ is equivalent to the length $S$ defined for RWI [Eq. (\ref{defS})],
and $R$ may be directly measured in an experiment.

\subsection{Three dimensions and localized source}
\label{fluid3d}

The model can be extended to three dimensions, where Eq. (\ref{laplacianh}) for the head leads to
\begin{equation}
\frac{\partial}{\partial r}\left(r^2 \frac{\partial h}{\partial r}\right)=0, \  \  \ h(a)=1, \ h(R)=0
\label{eqh3dpoint}
\end{equation}
with the same moving front condition as in the cylindrical case
\begin{equation}
- \left(\frac{\partial h}{\partial r} \right)_{r=R}=\frac{dR}{dt} .
\label{front3dpoint}
\end{equation}
The solution of Eq. (\ref{eqh3dpoint}) in $ a\le r \le R$ is
\begin{equation}
h=\frac{1}{r}\frac{aR}{R-a}-\frac{a}{R-a} .
\label{h3dpoint}
\end{equation}
Substituting Eq. (\ref{h3dpoint}) in Eq. (\ref{front3dpoint}), we obtain the equation for the
movement of the front as
\begin{equation}
\frac{dR}{dt}=\frac{a}{R^2-aR}, \  \  \ R(0)=a .
\label{eqR3dpointapprox}
\end{equation}
The solution for $R(t)$ is
\begin{equation}
\frac{R^3}{3a}-\frac{R^2}{2}+\frac{a^2}{6}=t .
\label{R3dpoint}
\end{equation}
In limit where the front displacement is much larger than the source radius, $R>>a$, we have
\begin{equation}
R\approx {\left( 3at\right)}^{1/3}  
\label{R3dpointapprox}
\end{equation}
and the infiltrated volume is
\begin{equation}
V\approx 4\pi at 
\label{V3dpointapprox}
\end{equation}

Here, the fluid infiltration shows the same linear time increase of the RWI in three dimensional
homogeneous media.
Note that the radius $R$, which may be experimentally measured, has a subdiffusive scaling,
which contrasts the rms displacement of RWI; again, $R$ is analogous to the length $S$ defined for RWI.

\subsection{Extension to fractal media}
\label{universality}

We simulated the fluid infiltration model in the $3^{\text{rd}}$ stage of construction of
the carpet with $b=3$ and $m=1$ and with a point source in an external boundary using
the same control volume finite difference methods of Ref. \protect\cite{Voller_wrr}.
Essentially, at each time step, this involves the numerical solution of
$\nabla (K \nabla h)=0$, with $K=1$ in the spaces and $K<<1$ in the obstacles.
In the current simulations, we use a  $3^5 \times 3^5$ mesh of square node centered control
volumes of size $\Delta =1/3^5$ and a time step of $\Delta t= 0.005$;
as fully explained in Ref. \protect\cite{Chen}, the accuracy of the calculation is independent 
of the time step size.

\begin{figure}[!ht]
\center
\includegraphics[clip,width=.5\textwidth,angle=0]{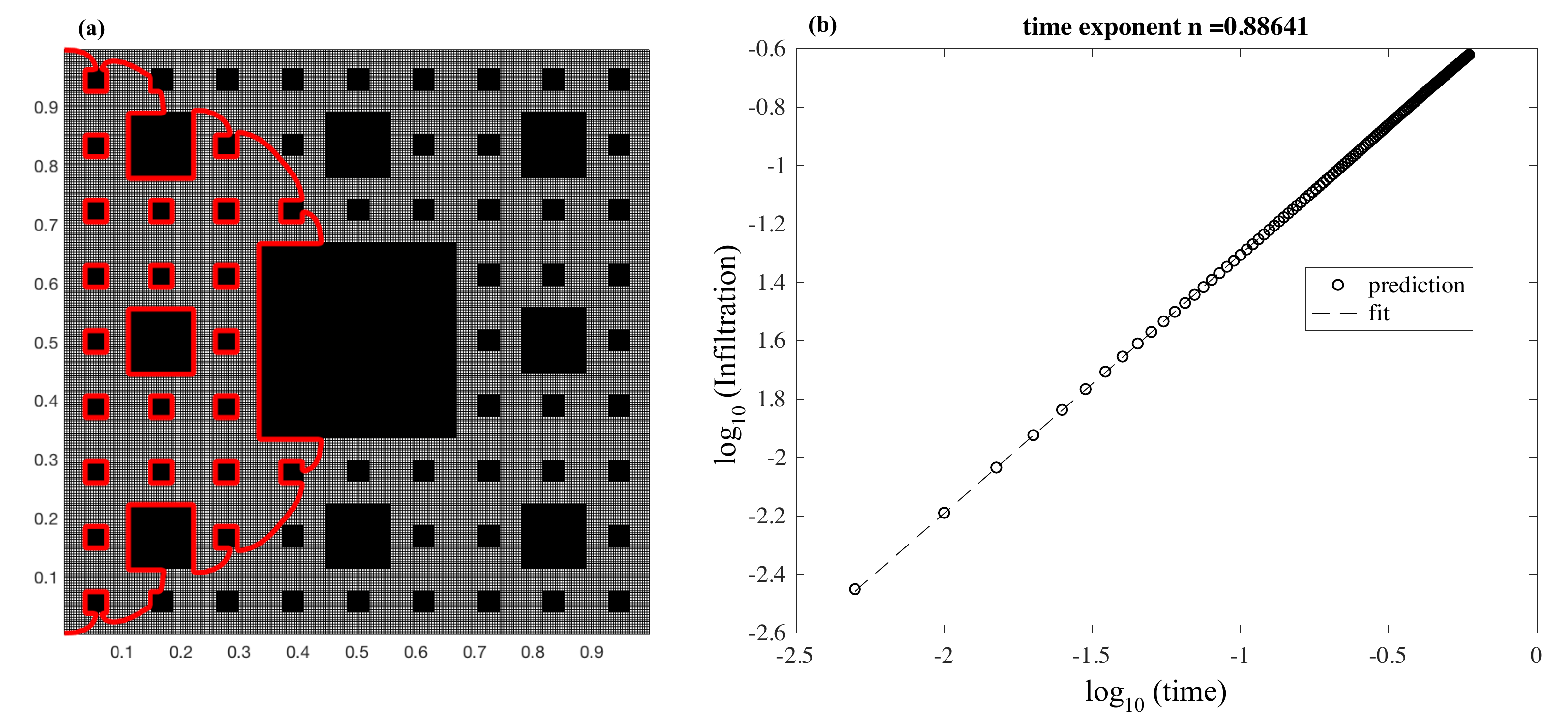}
\caption{Area filled with fluid as a function of time in the carpet with $b=3$ and $m=1$.
(a) The infiltration front at the point the fluid reaches the horizontal boundary.
(b) The log-log plot of the infiltration versus time. 
}
\label{VVadd2}
\end{figure}

In Fig. \ref{VVadd2}a, we show the infiltration front at the time in which the fluid reaches
the horizontal domain boundaries.
In Fig. \ref{VVadd2}b, we show the time evolution of the infiltrated area.
This area scales with an exponent close to the theoretical value $0.903$, which is also in
agreement with that of RWI in the same carpet (Sec. \ref{carpets}).

In this simulation, the dimension of the gate $a$ for the infiltration is 1/9 the size of the
smallest obstacle dimension in the carpet; we would expect values closer to the theory as the
relative size of the gate is reduced.
In fact, simulations on the geometry in Fig. \ref{VVadd2} with coarse grids and thus wider gates
show a quadratic dependence between the exponent $n$ and gate size;
in the continuous limit ($a\rightarrow 0$), this quadratic fit suggests an exponent $n=0.8994$, which differs less
than $0.5\%$ from the theoretical value.

\section{Conclusion}
\label{conclusion}

We studied a random walk infiltration model with small sources at the boundaries of homogeneous
and fractal media.
The model represents the penetration of a solute from a reservoir in contact with a porous medium by
a narrow hole, viz. the source.
A scaling approach proposes that the number of infiltrated particles increases in time
as the number of distinct sites visited by a single random walker in the same medium,
$N_D\left( t\right)$.
This is confirmed by numerical results obtained by direct simulations of the lattice model and
by integration of the diffusion equation that represents the model in the continuous limit.
In a Sierpinski carpet, the numerical integration of the diffusion equation provides an exponent
for the time evolution of the infiltration which is very close to the value of the
Alexander-Orbach conjecture for $N_D$, improving previous results obtained from random walk models.
In a Menger sponge where the fractal dimension is larger than the random walk dimension, the
infiltrated volume increases linearly in time, similarly to a homogeneous medium.

\begin{figure}[!ht]
\center
\includegraphics[clip,width=.35\textwidth,angle=0]{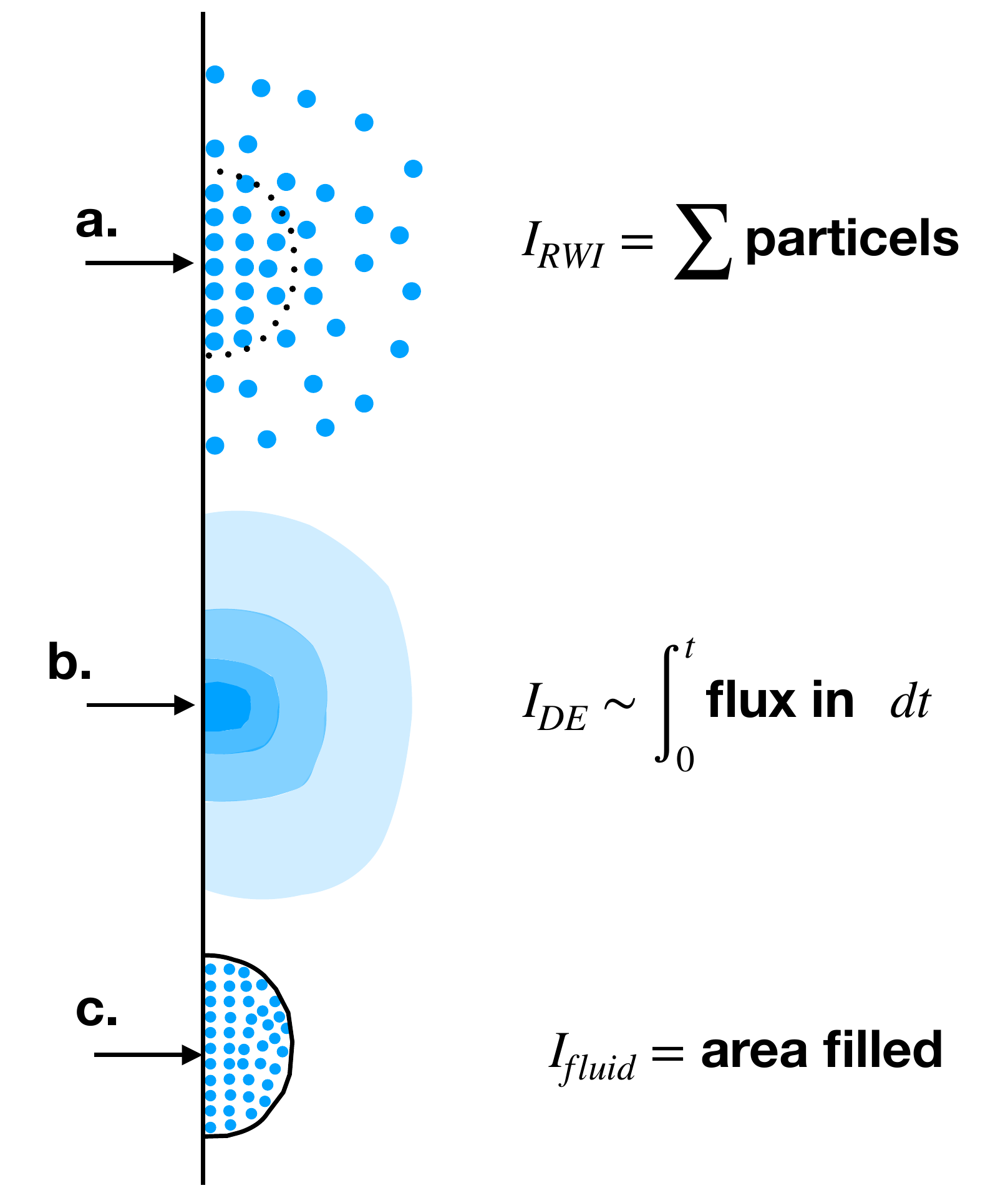}
\caption{Despite the different natures of the conservation of the RWI, diffusion equation, and
fluid infiltration processes, they all have identical scaling,
$I_{RWI}\sim I_{DE}\sim I_{fluid} \sim t^n$.
}
\label{VVadd3REV}
\end{figure}

We also studied a fluid infiltration model in the same geometry, with a constant pressure head
at the source and with the front displacement driven by the local gradient of that head.
Exact and numerical solutions in two and three dimensions and in Sierpinski carpets
show that the fluid infiltration model is in the same universality class of the random walk
infiltration.
The main difference between these models is the sharp front of fluid infiltration, while in RWI
the front is diffuse; in Fig. \ref{VVadd3REV}, these differences are highlighted, as well as the
differences between the random walk infiltration in a lattice and the solution of the
corresponding diffusion equation.
However, in the two models, the evolution of the fronts are driven by the gradient
of two analogous fields, which are the fluid head and the particle concentration.
Note that the equivalence between the two problems is observed in cases where random walks are
recurrent (carpets or two-dimensional media) and not recurrent (three-dimensional homogeneous media).
The scaling approach then sets a relation between the different infiltration
models from localized sources and the recurrence properties of random walks.

In fractal media where random walks are not recurrent, we show the failure of a previous exponent
relation for infiltration from flat boundaries \cite{Reis2016}.
This is the case of the Menger sponge studied here and of other fractals embedded in three dimensions
with the fractal dimension exceeding the random walk dimension (typically fractals with large
dimensions and weak subdiffusion).
In these cases, the normal infiltration scaling of three dimensional systems, $I\sim t$, is obtained,
which hides the fractality of the infiltrated medium.

As a final note, we recall that normal diffusion was already observed in several fractals,
i.e. $D_W=2$ \cite{burioni1994,forte2014,balankin2018}.
However, infiltration may be anomalous in those systems because the exponent
$n$ also depends on the dimension of the medium and on the dimension of the source,
as shown here for the case of small localized sources and in previous works for extended sources.


\begin{acknowledgments}

FDAAR acknowledges support by the Brazilian agencies CNPq (304766/2014-3) and FAPERJ 
(E-26/202941/2015) and thanks the hospitality of the Department of Civil, Environmental,
and Geo-Engineering of University of Minnesota, where part of this work was done.
VRV acknowledges support from the James L. Record Professorship.

\end{acknowledgments}


%

\end{document}